# Highly Variable Quasar Candidates Selected from 4XMM-DR13 with Machine Learning

Heng Wang[1], Yanli Ai[1], Yanxia Zhang[2], Yuming Fu[3,4], Wenfeng Wen[1], Liming Dou[5], Xue-Bing Wu[6,7], Xiangru Li[8], and Zhiying Huo[2]
[1] Shenzhen Key Laboratory of Ultraintense Laser and Advanced Material Technology, Center for Intense Laser Application Technology, and College of Engineering Physics, Shenzhen Technology University, Shenzhen 518118, People's Republic of China; aiyanli@sztu.edu.cn
[2] CAS Key Laboratory of Optical Astronomy, National Astronomical Observatories, Beijing 100101, People's Republic of China
[3] Leiden Observatory, Leiden University, Einsteinweg 55, 2333 CC Leiden, The Netherlands
[4] Kapteyn Astronomical Institute, University of Groningen, PO Box 800, 9700 AV Groningen, The Netherlands
[5] Department of Astronomy, School of Physics and Material Sciences, Guangzhou University, Guangzhou 510006, People's Republic of China
[6] Department of Astronomy, School of Physics, Peking University, Beijing 100871, People's Republic of China
[7] Kavli Institute for Astronomy and Astrophysics, Peking University, Beijing 100871, People's Republic of China
[8] School of Computer Science, South China Normal University, 510631 Guangzhou, People's Republic of China

## Abstract

We present a sample of 12 quasar candidates with highly variable soft X-ray emission, selected from the fourth XMM-Newton Serendipitous Source Catalog (4XMM-DR13), using random forest (RF). Optical to mid-IR photometric data for the 4XMM-DR13 sources were obtained by correlating the sample with the Sloan Digital Sky Survey (SDSS) DR18 photometric catalog and the AllWISE database. By further cross matching with known spectral catalogs from the SDSS and LAMOST surveys, we compiled a training data set containing stars, galaxies, and quasars. The RF algorithm was trained to classify the XMM–Wide-field Infrared Survey Explorer–SDSS sample. We then refined the quasar candidate selection by applying Gaia proper motion data to eliminate stellar contaminants. As a result, 52,486 quasar candidates were classified, with 8410 of them matching known quasars in SIMBAD. The quasar candidates exhibit systematically lower X-ray fluxes compared to quasars in the training set, suggesting that the classifier is effective in identifying fainter quasars. From this quasar candidate sample, we constructed a subset of 12 sources that have shown variations in their soft X-ray flux by a factor of 10 over ∼20 yr in the XMM-Newton survey. These highly variable quasar candidates extend the quasar sample characterized by extreme soft X-ray variability to the optically faint end, with magnitudes around $r \sim 22$. Notably, none of these 12 sources were detected in ROSAT observations. Given the flux sensitivity of ROSAT, the result indicates that quasars exhibiting more than 2 orders of magnitude of variation are extremely rare.

*Unified Astronomy Thesaurus concepts:* Quasars (1319); X-ray quasars (1821)

*Materials only available in the online version of record:* machine-readable table

## 1. Introduction

The X-ray emission of active galactic nuclei (AGNs) is considered to be largely generated in the accretion-disk corona via inverse Compton scattering of the optical/UV radiation (e.g., R. Sunyaev & L. Titarchuk 1980; F. Yuan & R. Narayan 2014). Variability of the X-ray emission of AGNs is a useful probe of the underlying nature of the corona and innermost accretion flow encircling the supermassive black hole. X-ray variability from AGNs has been observed over a wide range of timescales from hours to years, with larger variability amplitudes observed on longer timescales. The observed red-noise-like X-ray power spectrum density in AGNs can usually be described by a broken power law, with a slope of −2 at high frequencies and a flatter slope of −1 at lower frequencies (e.g., P. Uttley et al. 2002; O. Gonzalez-Martin & S. Vaughan 2012). The characteristic timescales can be physically linked to AGNs and black hole X-ray binaries, indicating that the accretion process of supermassive black holes in AGNs is similar to that of smaller black holes (I. M. McHardy et al. 2006).

The X-ray flux variability amplitude intrinsic to the accretion disk and corona of AGNs typically does not exceed a factor of ∼2 (e.g., R. R. Gibson & W. Brandt 2012; R. Middei et al. 2017; J. D. Timlin et al. 2020). Extreme X-ray variations with amplitudes exceeding a factor of 10 or more in AGNs are still rare. Strong X-ray variability events have been observed in a few typical type 1 AGNs (e.g., M. Mehdipour et al. 2021; Y. Wang et al. 2022), which were attributed to changes in the column density of the dust-free obscuring material along the observer's line of sight. High-amplitude changes in X-ray flux were also found in a few cases of changing-look AGNs, associated with significant X-ray spectral variability (e.g., D. Grupe et al. 2015; M. Krumpe et al. 2017; Y. Ai et al. 2020; Z. Liu et al. 2020; A. Jana et al. 2021; Q. Yang et al. 2023). This extreme X-ray variation can be interpreted in the scheme of destruction and recreation of the inner accretion disk and corona. In the case of AGN 1ES 1927+654, the extreme X-ray variation can also be explained between the interactions of the accretion disk and debris from a tidally disrupted star (C. Ricci et al. 2020; X. Cao et al. 2023).

Strong and sometimes rapid X-ray variability has been observed in some narrow-line Seyfert 1 galaxies (NLS1s), which are considered to have high or even super-Eddington accretion rates (e.g., J. Reeves & V. Braito 2019; T. Boller et al. 2021; M. L. Parker et al. 2021; C. Jin et al. 2023). Powerful disk winds launched via radiation pressure are generally expected in these systems (e.g., Y.-F. Jiang et al. 2019; H. Yang & F. Yuan 2024), and one possible explanation for the observed variability is fluctuating X-ray obscuration due to clumpy accretion-disk winds. Extreme X-ray variations in higher luminosity quasars appear to be much rarer. Recently, a few such objects have been confirmed to exhibit extreme X-ray variations (e.g., N. Strotjohann et al. 2016; Z. Liu et al. 2020; Q. Ni et al. 2020; S. Wang et al. 2024). Most of these quasars are considered to have high or even super-Eddington accretion rates, similar to NLS1s. The strong X-ray variability detected in these high-luminosity quasars is also likely to be related to dust-free obscuration from accretion-disk winds.

Sample studies of X-ray highly variable AGNs have been conducted with the ROSAT All-Sky Survey (RASS), XMM Serendipitous Source Catalog (XMMSL2)/XMM-Newton slew survey, Spektrum Roentgen Gamma (SRG)/eROSITA all-sky survey (D. Bi et al. 2015; N. Strotjohann et al. 2016; P. Medvedev et al. 2022). Most of them are bright AGNs with soft X-ray flux greater than $10^{-14}$ erg s$^{-1}$ cm$^{-2}$. The analysis in J. D. Timlin et al. (2020), using serendipitous Chandra X-ray observations, indicates that extreme intrinsic X-ray variations are rare in radio-quiet quasars, with a maximum occurrence rate of <2.4% of observations. The quasars in the investigation are optically bright, with $m_i \leqslant 20.2$.

XMM-Newton has been surveying the sky in the X-ray band for 20 yr. While this sky survey produces enormous amounts of data covering extensive regions of the sky, approximately 80% of sources remain unclassified in 4XMM-DR13 (N. A. Webb et al. 2020). Machine learning algorithms have proven efficient in automatically classifying X-ray sources, as demonstrated by the classification of 4XMM-DR9 sources in Y. Zhang et al. (2021). In the soft energy band (0.2–2 keV), the median flux in the 4XMM-DR13 catalog is ~5.2 × $10^{-15}$ erg s$^{-1}$ cm$^{-2}$, which is about 2 orders of magnitude deeper than the RASS/XMMSL2. Automated classification of X-ray sources is becoming increasingly valuable for identifying rare objects, such as tidal disruption events, changing-look AGNs, and ultraluminous X-ray sources.

In this paper, we first classify 4XMM-DR13 sources using a random forest (RF) approach with multiwavelength data and then perform a selection of highly variable X-ray sources among the classified quasar candidates. This paper is organized as follows. We describe the correlated 4XMM-DR13, Sloan Digital Sky Survey (SDSS), and AllWISE data in Section 2. In Section 3 we give a detailed explanation of the RF approach and the classification results. We present the process of selecting highly variable quasar candidates in Section 4. Finally, we discuss our results in the context of highly variable AGN populations in Section 5, and summarize in Section 6.

## 2. The Data

### 2.1. 4XMM-DR13 and Multiwavelength Data

The 4XMM-DR13 catalog contains source detections drawn from 13,243 XMM-Newton European Photon Imaging Camera (EPIC) observations (N. A. Webb et al. 2020), covering an energy interval from 0.2 to 12 keV. These observations were made between 2000 February 3 and 2022 December 31. All data sets, including 983,948 detections from 656,997 unique sources, were publicly available by 2022 December 31. We select reliable pointlike sources that have no detected extent and are considered to be clean (SUM_FLAG < 2) in at least one detection. Furthermore, for the sources, we only keep the observations in which the object has a detection likelihood >10 and was detected with more than 5 counts. There are 433,530 sources left with 648,565 detections. The X-ray properties from 4XMM-DR13 for these selected sources include energy fluxes and the hardness ratio derived from different energy bands.

**Table 1**
The Number of Known Samples

| Class | No. | Proportion |
|---|---|---|
| Galaxy | 2875 | 12.23% |
| Star | 3800 | 16.17% |
| QSO | 16,826 | 71.60% |
| Total no. | 23,501 | 100.00% |

We cross match the 4XMM-DR13 catalog with the Sloan Digital Sky Survey (SDSS; D. G. York et al. 2000) DR18 and AllWISE (E. F. Schlafly et al. 2019) databases to obtain multiwavelength properties of X-ray sources. The matching radius is 6″ for SDSS and 8″ for AllWISE, as proposed by Y. Zhang et al. (2021). We use the software TOPCAT (M. B. Taylor 2005) and the website CasJobs[9] for cross matching the data. For the matched sources in SDSS, we first filter the data with a photometry flag to get a cleaned sample, and we include only those with magnitudes below the limiting magnitudes in each band and with corresponding magnitude errors no larger than 0.2. The fluxes for sources brighter than the nominal saturation limits (W1 < 8, W2 < 7) observed during the NEOWISE Post-Cryo phase exhibit larger uncertainties and more scatter due to the measurements being derived from the 4-Band Cryo data used in the Wide-field Infrared Survey Explorer (WISE) All-Sky Release Catalog. For the matched sources in AllWISE, the data are filtered to satisfy 8 < W1 < 17.7 and 7 < W2 < 17.5. All the SDSS *ugriz* magnitudes have been corrected from the extinction using a map from E. F. Schlafly et al. (2019). Throughout the paper, the SDSS magnitudes are given in Vega magnitudes. In the SDSS and AllWISE databases, all sources corresponding to the XMM objects are retained, regardless of whether there are missing magnitudes. Thus 100,183 sources are selected with qualified data from XMM-Newton, SDSS, and AllWISE. In the following, these sources are designated as the XMM-WISE-SDSS sample.

We then cross match these sources with the objects identified spectroscopically by SDSS DR18 and the Large Sky Area Multi-object Fiber Spectroscopic Telescope (LAMOST; X.-Q. Cui et al. 2012; A. L. Luo et al. 2015) DR9, using a 5″ radius. In the process of matching with the SDSS DR18 SpecObj database, we select good spectra with no identified problems by setting the "zWarning = 0" in CasJobs. The sources were identified as stars, galaxies, or QSOs by the SDSS DR18 and LAMOST automated classification pipelines using template fitting. We further cross match the XMM-WISE-SDSS sample with the SDSS quasar catalog from Data Release 16 (DR16Q). The sources (∼0.5%) with different classification between SDSS DR18/LAMOST and DR16Q were labeled as QSOs. Finally, the known sample included 3800 stars, 2875 galaxies, and 16,826 quasars with data from the X-ray, optical, and infrared bands, as shown in Table 1.

We select the features $\log(f_x)$, $hr1$, $hr2$, $hr3$, $hr4$, $r$, W1, W2, $u - g$, $g - r$, $r - i$, $i - z$, $z -$ W1, W1 $-$ W2, and $\log(f_x/f_r)$ from the known star, galaxy, and quasar samples used in this

---
[9] https://skyserver.sdss.org/casjobs/

Table 2
Definition, Catalog, and Wave Band of Parameters for the Training Sample

| Parameter | Definition | Catalog | Wave Band |
|---|---|---|---|
| $SC\_HR1$ | Hardness ratio 1 definition: $hr1 = (B - A)/(B + A)$, where A = count rate in energy band 0.2–0.5 keV, B = count rate in energy band 0.5–1 keV | XMM | X-ray band |
| $SC\_HR2$ | Hardness ratio 2 definition: $hr2 = (C - B)/(C + B)$, where B = count rate in energy band 0.5–1 keV, C = count rate in energy band 1–2 keV | XMM | X-ray band |
| $SC\_HR3$ | Hardness ratio 3 definition: $hr2 = (D - C)/(D + C)$, where C = count rate in energy band 1–2 keV, D = count rate in energy band 2–4.5 keV | XMM | X-ray band |
| $SC\_HR4$ | Hardness ratio 4 definition: $hr2 = (E - D)/(E + D)$, where D = count rate in energy band 2–4.5 keV, E = count rate in energy band 4.5–12 keV | XMM | X-ray band |
| $\log(f_x)$ | X-ray flux in 0.2–12 keV | XMM | X-ray band |
| $\log(f_x/f_r)$ | X-ray-to-optical flux ratio | XMM, SDSS | Optical and X-ray bands |
| W1 | W1 magnitude | AllWISE | Infrared band |
| W2 | W2 magnitude | AllWISE | Infrared band |
| r | r magnitude | SDSS | Optical band |
| u − g | u − g color | SDSS | Optical band |
| g − r | g − r color | SDSS | Optical band |
| r − i | r − i color | SDSS | Optical band |
| i − z | i − z color | SDSS | Optical band |
| z − W1 | z − W1 color | AllWISE, SDSS | Infrared and optical bands |
| W1 − W2 | W1 − W2 color | AllWISE | Infrared band |

Table 3
The Performance for Four Different Settings

| Metric Condition | Precision (%) | | | Recall (%) | | | F1-score (%) | | | TA (%) | TF (%) |
|---|---|---|---|---|---|---|---|---|---|---|---|
| | Galaxy | QSO | Star | Galaxy | QSO | Star | Galaxy | QSO | Star | | |
| I | 87.06 | 95.36 | 97.58 | 79.45 | 98.53 | 89.85 | 83.08 | 96.92 | 93.56 | 94.75 | 91.18 |
| II | 89.36 | 95.53 | 97.96 | 81.39 | 98.76 | 89.64 | 85.19 | 97.12 | 93.62 | 95.19 | 91.98 |
| III | 93.97 | 93.59 | 99.06 | 96.32 | 95.38 | 94.60 | 95.13 | 94.48 | 96.78 | 95.43 | 95.46 |
| IV | 94.97 | 93.95 | 99.41 | 97.20 | 96.29 | 94.60 | 96.07 | 95.10 | 96.95 | 96.03 | 96.04 |

**Note.** Condition I represents default model parameters without SMOTE. Condition II refers to optimal model parameters without SMOTE. Condition III denotes default model parameters with SMOTE. Condition IV corresponds to optimal model parameters with SMOTE. TA stands for total accuracy, while TF stands for total F1-score.

study. The selected features are described in Table 2. These attributes contribute to the classification to varying extents, as discussed in Y. Zhang et al. (2021).

### 2.2. Imbalanced Data

As shown in Table 1, our data set is imbalanced, with QSOs comprising 71.60% of the total sources, significantly outnumbering the other two classes. This imbalance can severely impact classifier performance, leading it to favor the majority class when classifying unidentified sources. Several methods can mitigate this issue. One straightforward approach is to assign weights to the source classes, thereby penalizing the algorithm more for misclassifying sources as the majority class. However, increasing the weight of the minority classes in this way could cause the model to overemphasize these samples, potentially resulting in overfitting to the training data (TD) and reducing generalization performance on new, unidentified data.

Another way to solve the imbalance problem is the use of an oversampling algorithm, such as the Synthetic Minority Oversampling Technique[10] (SMOTE; A. Luque et al. 2019). For each minority class sample, SMOTE uses the K nearest neighbor (KNN) algorithm to find its K nearest neighbors, which should also belong to the minority class. One neighbor is randomly selected from its K nearest neighbors (K = 5 is adopted in this study). For each feature, the difference between the selected neighbor and the current minority class sample is calculated. A random number λ (ranging from 0 to 1) is generated, and this difference is multiplied by λ. The resulting product is then added to the features of the original minority class sample, creating a new synthetic sample. This process is repeated until the data set is balanced or until the desired number of synthetic samples is generated.

To evaluate the effectiveness of SMOTE in this problem, this work conducted four experiments: one with default model parameters without SMOTE (Condition I), one with optimal model parameters without SMOTE (Condition II), one with default model parameters with SMOTE (Condition III), and one with optimal model parameters with SMOTE (Condition IV). The experimental results are shown in Table 3. For four different conditions, the total F1-score (see the definition of F1-score in Section 3.2) is 91.18%, 91.98%, 95.46%, and 96.04%, respectively, and the total accuracy is 94.75%, 95.19%, 95.43%, and 96.03%, respectively. In terms of total F1-score and total accuracy, the best performance is achieved with optimal model parameters and SMOTE. In this condition, the performance on stars and galaxies improves, which is beneficial to reduce the contamination of stars and galaxies in targeting the quasar task. These results indicate that the SMOTE algorithm enhances the diversity of the minority class by generating new synthetic samples, which helps mitigate the risk of model overfitting. After balancing the categories, the classifier's ability to identify minority classes is significantly improved, leading to better overall classification performance. Therefore, this work uses the SMOTE algorithm to deal with the imbalanced samples.

---
[10] https://imbalanced-learn.org/

Additionally, missing data can occur due to insufficient survey depth, complex weather conditions, or lack of emission in specific bands. To address the impact of missing data on the SMOTE algorithm, we assign a placeholder value of −9999 to these missing entries.

## 3. Method

The primary method employed in this study to identify quasar candidates is the RF algorithm. The performance of the classifier is evaluated using metrics such as precision, recall, and accuracy. Additionally, we use cross validation to assess the robustness and generalizability of the classifier.

### 3.1. Random Forest

RF is an ensemble learning algorithm based on decision trees and bagging techniques. It mitigates overfitting through randomness in two dimensions: by randomly selecting subsets of training samples and by each decision tree using a different subset of features. This approach improves the model's robustness and reduces sensitivity to missing values. Due to its strong performance, RF is widely used in astronomical data processing. The method was initially introduced and termed "random decision forests" by T. K. Ho (1995) and later refined and renamed "random forests" by L. Breiman (2001). In this study, we construct our classifier based on the RF module in scikit-learn[11] (F. Pedregosa et al. 2011).

### 3.2. Evaluation Metrics

Accuracy, precision, recall, and F1-score are fundamental metrics for evaluating classification models from multiple perspectives. These metrics are derived from the counts of true positives (TPs), false positives (FPs), true negatives (TNs), and false negatives (FNs), each providing different insights into the model's performance. Specifically, accuracy is given by

$$\text{Accuracy} = \frac{\text{TP} + \text{TN}}{\text{TP} + \text{TN} + \text{FP} + \text{FN}}, \quad (1)$$

which stands for how well quasars and nonquasars are identified.

Precision is given by

$$\text{Precision} = \frac{\text{TP}}{\text{TP} + \text{FP}}. \quad (2)$$

Taking quasars as an example, precision means the percentage of quasars in the test set that can be correctly picked out among those classified as quasars.

---
[11] https://scikit-learn.org/stable/index.html

Recall is used to assess how many quasars are selected in the true quasars, which is defined as

$$\text{Recall} = \frac{\text{TP}}{\text{TP} + \text{FN}}. \quad (3)$$

F1-score is a metric commonly used in binary classification tasks to assess the performance of the classification models by considering both precision and recall. It is the harmonic mean of precision and recall, providing a balance between these two metrics:

$$\text{F1} - \text{score} = 2 \times \frac{\text{Precision} \times \text{Recall}}{\text{Precision} + \text{Recall}}. \quad (4)$$

### 3.3. K-fold Cross Validation

$K$-fold cross validation is a widely used technique for evaluating model performance and selecting optimal hyperparameters. In this method, the data set is partitioned into $K$ equal-sized subsets. For each iteration, $K - 1$ subsets are used to train the model, while the remaining subset is used for validation. This process is repeated $K$ times, with each subset serving as the validation set exactly once. The final performance metric is obtained by averaging the results from all $K$ iterations.

### 3.4. Hyperparameter Adjustment

The RF model has some adjusted hyperparameters. By adjusting these hyperparameters, we can find the best model parameters for the current problem to improve the performance of the model. The three most important parameters are the number of decision trees generated by the RF algorithm (*n_estimators*), the number of features to be considered when splitting nodes in each decision tree (*max_features*), and the maximum depth of each decision tree (*max_depth*). We estimate the optimal parameter configuration using a grid search algorithm with 5-fold cross validation. The optimal value is $\log_2 15$ for *max_features* and 151 for *n_estimators*. Based on the results, we do not impose an upper limit on *max_depth*.

## 4. Result

### 4.1. Additional Filtering with Gaia Proper Motions

The flowchart to classify 4XMM-DR13 sources is described in Figure 1. We classify the 4XMM-DR13 sources into 52,486 quasars, 15,340 galaxies, and 8856 stars with the RF algorithm. The classification outcomes are presented in Table 4. In Figure 2, we show the distribution of the classified quasar, galaxy, and star candidates in the optical/midinfrared color–color diagram. Quasars and galaxies are typically clustered around regions with the highest densities in the two-dimensional color spaces, which results in smooth contours in the diagrams. By contrast, stars are largely distributed on narrow stripes in color–color diagrams. The locations of quasar candidates in the color–color diagram are consistent with the results in the literature (X.-B. Wu et al. 2012; Y. Zhang et al. 2021; Y. Fu et al. 2024).

In order to remove stellar contaminants from quasar candidates, we then apply an additional cut based on Gaia's proper motion. We adopt the probabilistic cut algorithm introduced in Y. Fu et al. (2021, 2024), which accounts for the uncertainties in proper-motion measurement from Gaia. In this algorithm, the probability density of zero proper $f_{\text{PM0}}$ for a source is defined based on the bivariate normal distribution of

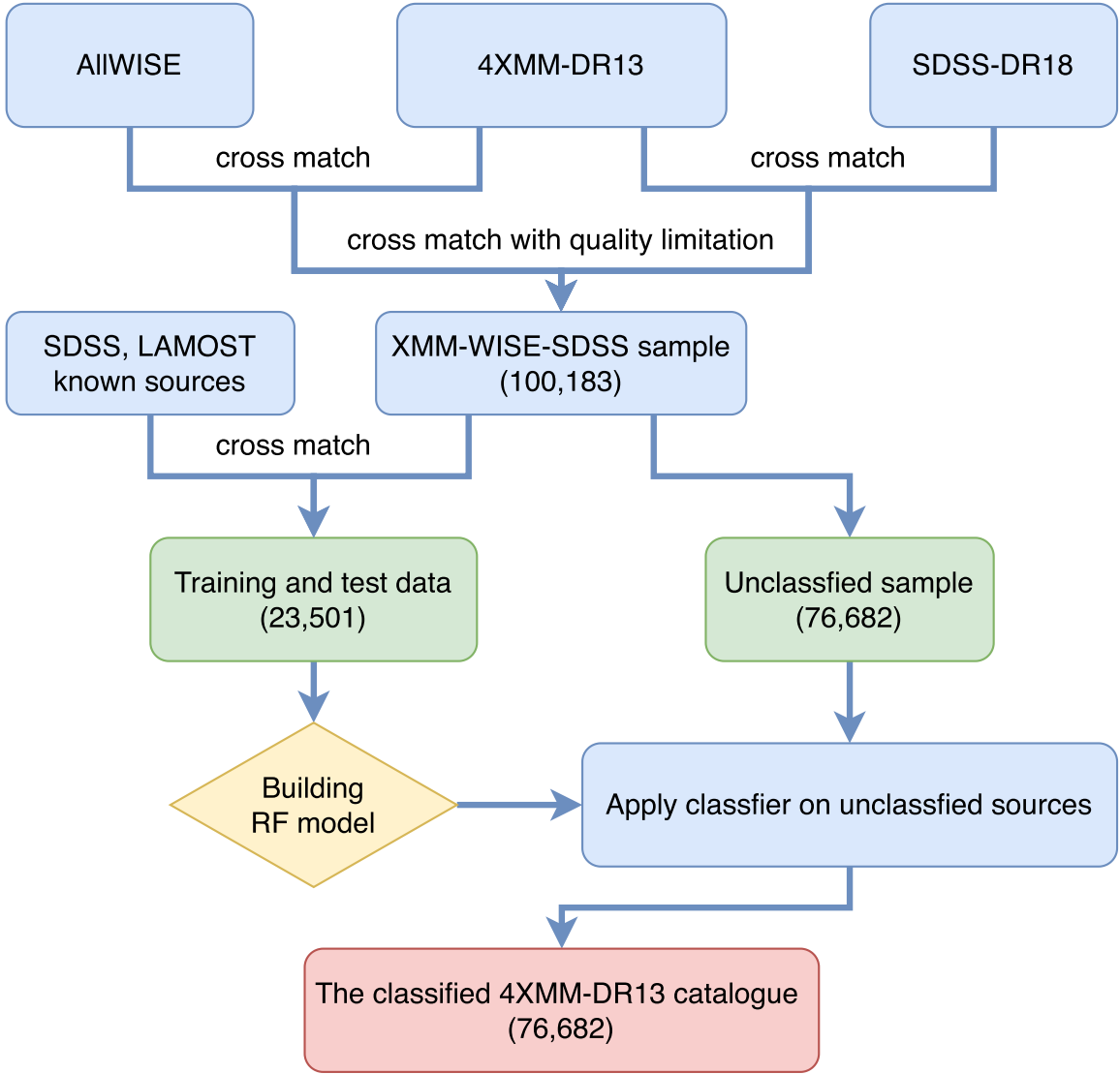

**Figure 1.** The classification flowchart for the 4XMM-DR13 sources.

proper-motion measurements as follows:

$$f_{\rm PM0} = \frac{1}{2\pi\sigma_x\sigma_y\sqrt{1-\rho^2}}$$
$$\times \exp\left\{-\frac{1}{2(1-\rho^2)}\left[\left(\frac{x}{\sigma_x}\right)^2 - \frac{2\rho xy}{\sigma_x\sigma_y} + \left(\frac{y}{\sigma_y}\right)^2\right]\right\}, \tag{5}$$

where $x$ represents pmra, $y$ represents pmdec, $\rho$ denotes pmra_pmdec_corr (correlation coefficient between pmra and pmdec), and $\sigma_x$ and $\sigma_y$ are the proper-motion uncertainties. Following Y. Fu et al. (2024), we apply a cut of $\log(f_{\rm PM0}) \geqslant -4$ to exclude potential stars while retaining most quasars.

We cross match the classified 4XMM-DR13 quasar candidates with Gaia Data Release 3 (DR3). The crossing radius is set to 5″, which accounts for most of the position uncertainties (e.g., X.-j. Xu et al. 2022). A total of 15,414 sources are matched, of which 872 objects are potentially stellar contaminants with $\log(f_{\rm PM0}) < -4$. The result indicates the high precision of our quasar classification. These potentially stellar objects are removed from the quasar candidates, leaving 51,614 quasar candidates. We then correlate these quasar candidates with SIMBAD using a matching radius of 5″, identifying 8,410 quasars among 13,485 counterparts. Finally, we classify 38,129 unidentified 4XMM-DR13 sources as quasar candidates. In the following section, we describe the selection of X-ray highly variable sources from the classified quasar candidates. To assess the robustness of our approach, the same selection algorithms are also applied to the known quasars.

### 4.2. Quasar Candidates with Extreme X-Ray Variability

In this paper we focus on selecting the X-ray highly variable, radio-quiet quasars as their variability mechanisms are quite different from those of radio-loud quasars, which exhibit a highly collimated emission component. To obtain a sample devoid of jet emission, we cross match the classified quasar candidates with the Faint Images of the Radio Sky at Twenty centimeters (FIRST) survey using a matching radius of 5″. Instead of using the radio-loudness parameter to exclude radio-loud quasars, we simply remove their FIRST counterparts from the sample, leaving 37,818 quasar candidates. For the selection

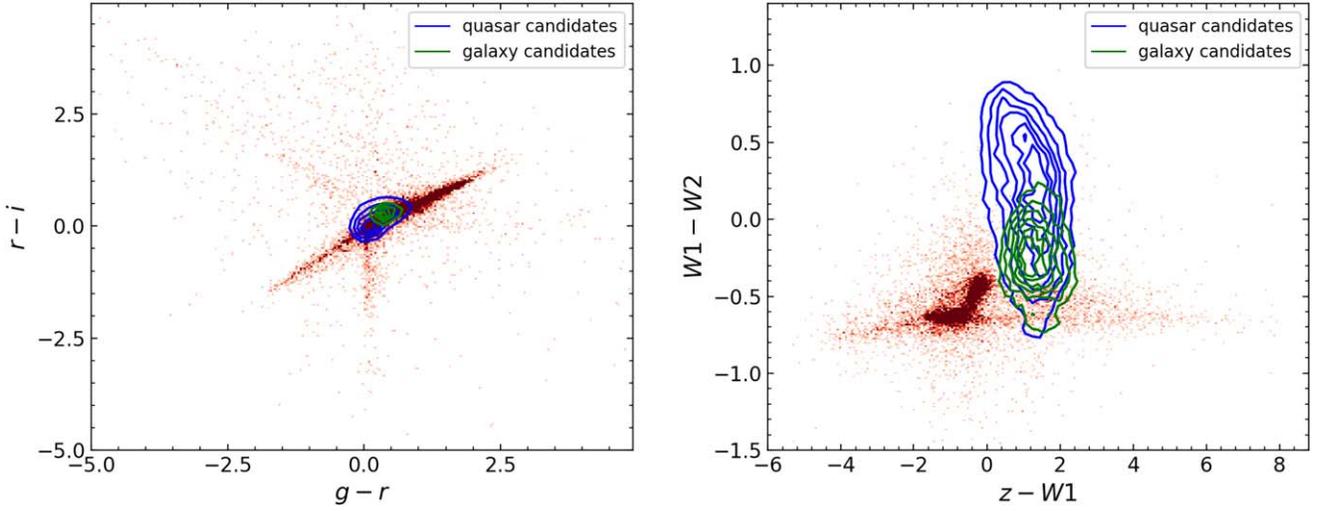

**Figure 2.** Color–color diagrams of quasar candidates (blue contours), galaxy candidates (green contours), and star candidates (red-shaded density plots).

**Table 4**
The Classification Result of the XMM-WISE-SDSS Sample by Random Forest

| SRCID | R.A. | Decl. | Class_RF | P_GALAXY | P_QSO | P_STAR |
|---|---|---|---|---|---|---|
| 207427402010003 | 207.739511 | −3.602388 | QSO | 0.15894 | 0.741722 | 0.099338 |
| 206337701010032 | 40.79156 | 32.291503 | STAR | 0.152318 | 0.019868 | 0.827815 |
| 207622902010059 | 34.483102 | −0.753867 | GALAXY | 0.715232 | 0.18543 | 0.099338 |
| 201094614015002 | 153.85554 | 9.243593 | QSO | 0.291391 | 0.642384 | 0.066225 |
| 206553438370016 | 35.702477 | −6.445397 | QSO | 0.006623 | 0.887417 | 0.10596 |
| 202000201010021 | 189.55514 | 17.655152 | QSO | 0.059603 | 0.907285 | 0.033113 |
| 204049601010024 | 36.614497 | −3.818808 | QSO | 0.357616 | 0.582781 | 0.059603 |
| 201034608010002 | 244.062116 | 12.231999 | QSO | 0.019868 | 0.94702 | 0.033113 |
| 206562002010052 | 321.198259 | 20.0555 | QSO | 0.192053 | 0.761589 | 0.046358 |
| 201112824010029 | 205.625455 | 0.59069 | QSO | 0.10596 | 0.887417 | 0.006623 |
| 201122402010052 | 230.650038 | 27.55744 | QSO | 0.15894 | 0.701987 | 0.139073 |
| 206703505010074 | 247.033204 | 42.50509 | QSO | 0.178808 | 0.761589 | 0.059603 |
| 206553401580011 | 135.511154 | −2.40823 | QSO | 0.02649 | 0.92053 | 0.05298 |
| 202047904010004 | 9.947335 | 41.529773 | QSO | 0.006623 | 0.960265 | 0.033113 |
| 207603402010092 | 13.722353 | −1.523961 | QSO | 0.119205 | 0.794702 | 0.086093 |
| 205545008010012 | 51.15235 | 40.721397 | QSO | 0.192053 | 0.735099 | 0.072848 |
| 202006501010032 | 189.371394 | 13.197505 | GALAXY | 0.569536 | 0.403974 | 0.02649 |
| 200335409010041 | 248.265588 | 37.606769 | GALAXY | 0.523179 | 0.430464 | 0.046358 |
| 200936301010072 | 40.071859 | −8.2434 | QSO | 0.039735 | 0.907285 | 0.05298 |
| 204064204010030 | 259.636705 | 58.877744 | QSO | 0.251656 | 0.523179 | 0.225166 |

**Note.** P_GALAXY, P_QSO, and P_STAR are the classification probabilities of galaxies, QSOs, and stars, respectively. The full table is also available at https://github.com/plunck/XMMDR13-cand/blob/main/XMMDR13_cand.csv.

(This table is available in its entirety in machine-readable form in the online article.)

of X-ray variable quasars, we include only objects with at least two XMM-Newton flux measurements. The resulting sample consists of 11,232 quasars with 33,808 XMM-Newton observations. Among these quasars, approximately 40% are detected in more than two epochs.

We utilize the flux ratio in the soft band between XMM-Newton epochs to select highly variable quasars. The soft X-ray (0.2–2.0 keV) flux is calculated as the sum of the EPIC fluxes listed in the 4XMM-DR13 catalog for bands 1, 2, and 3 (0.2–0.5, 0.5–1, and 1–2 keV, respectively). These observed fluxes are derived from the count rates, assuming an absorbed power-law spectral model with a photon index $\Gamma = 1.7$ and a Galactic absorption column density of $N_H = 3 \times 10^{20}\,\mathrm{cm}^{-2}$ (N. A. Webb et al. 2020). For these selected extreme variable quasars, we can neglect the inaccuracies introduced in the flux ratios by use of this fixed spectral slope (R. Saxton et al. 2011; D. Li et al. 2022).

We select the candidates with a soft X-ray flux ratio between XMM-Newton epochs greater than a factor of 10. Considering the larger uncertainties in the measured flux of the relatively fainter sources, we reject the candidates with fluxes fainter than $5 \times 10^{-15}\,\mathrm{erg\,cm}^{-2}\,\mathrm{s}^{-1}$. This flux threshold corresponds to the median value in the soft X-ray band from the 4XMM-DR13 catalog (N. A. Webb et al. 2020). At this stage, 103 highly variable quasar candidates are selected. We then cross match these sources with the NASA Extragalactic Database (NED) to exclude objects that have been spectroscopically classified. This results in the removal of 25 objects that are identified as

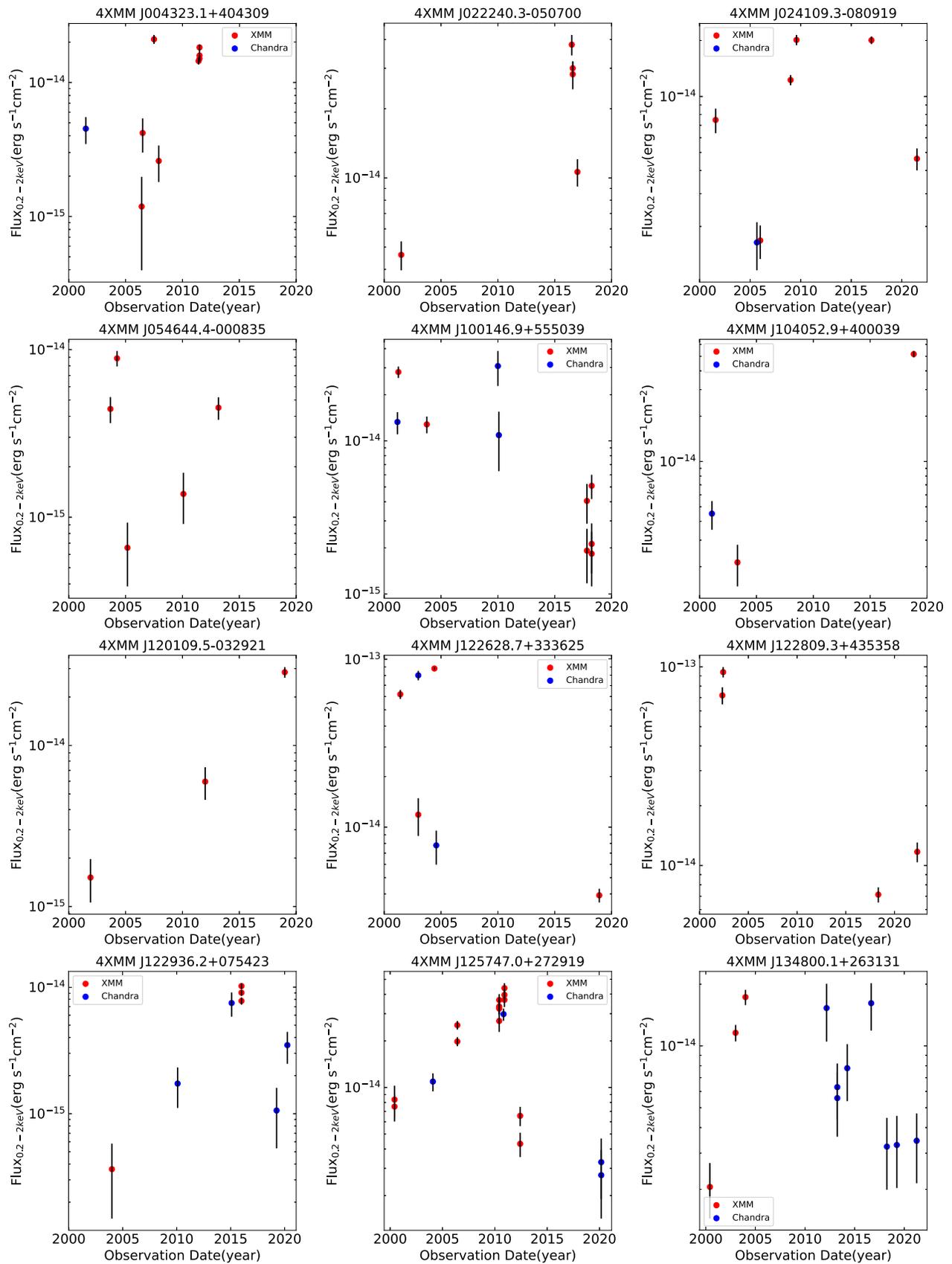

**Figure 3.** Light curves of the 12 quasar candidates with highly variable X-ray emission. All error bars are $3\sigma$.

**Table 5**
The Quasar Sample with Highly Variable X-Ray Soft Flux

| 4XMM Name | Photo_z[a] | XMM-Newton Flux[b] | Obs. Time | Chandra Flux[c] | Obs. Time |
|---|---|---|---|---|---|
| 4XMM J004323.1+404309 | 0.37 | 2.10 | 2007-07-22 | 0.45 | 2001-07-24 |
| … | … | 1.59 | 2011-07-13 | … | … |
| … | … | 0.12 | 2006-06-30 | … | … |
| … | … | 0.26 | 2007-12-29 | … | … |
| … | … | 1.45 | 2011-06-27 | … | … |
| … | … | 0.42 | 2006-07-08 | … | … |
| … | … | 1.51 | 2011-07-05 | … | … |
| … | … | 1.83 | 2011-07-07 | … | … |
| 4XMM J022240.3−050700 | … | 3.80 | 2016-07-08 | … | … |
| … | … | 1.06 | 2017-01-01 | … | … |
| … | … | 0.46 | 2001-07-04 | … | … |
| … | … | 2.83 | 2016-08-14 | … | … |
| … | … | 3.00 | 2016-08-13 | … | … |
| 4XMM J024109.3−080919 | 0.70 | 1.22 | 2009-01-14 | 0.16 | 2005-09-18 |
| … | … | 0.46 | 2021-07-16 | … | … |
| … | … | 0.75 | 2001-08-15 | … | … |
| … | … | 0.17 | 2006-01-12 | … | … |
| … | … | 2.01 | 2017-01-17 | … | … |
| … | … | 2.01 | 2009-08-12 | … | … |
| 4XMM J054644.4−000835 | … | 0.14 | 2010-02-28 | … | … |
| … | … | 0.89 | 2004-04-03 | … | … |
| … | … | 0.45 | 2013-03-09 | … | … |
| … | … | 0.44 | 2003-09-03 | … | … |
| … | … | 0.07 | 2005-03-24 | … | … |
| 4XMM J100146.9+555039 | … | 0.19 | 2017-11-01 | 1.09 | 2010-02-01 |
| … | … | 0.18 | 2018-04-17 | 3.08 | 2010-01-17 |
| … | … | 0.51 | 2018-04-23 | 1.33 | 2001-03-07 |
| … | … | 2.81 | 2001-04-13 | … | … |
| … | … | 0.41 | 2017-11-27 | … | … |
| … | … | 0.21 | 2018-04-21 | … | … |
| … | … | 1.28 | 2003-10-14 | … | … |
| 4XMM J104052.9+400039 | … | 5.17 | 2018-11-13 | 0.45 | 2001-02-04 |
| … | … | 0.21 | 2003-05-24 | … | … |
| 4XMM J120109.5−032921 | 0.70 | 0.60 | 2012-01-06 | … | … |
| … | … | 2.85 | 2019-01-04 | … | … |
| … | … | 0.15 | 2001-12-11 | … | … |
| 4XMM J122628.7+333625 | … | 1.19 | 2003-01-03 | 8.02 | 2003-01-27 |
| … | … | 8.81 | 2004-06-02 | 0.78 | 2004-08-07 |
| … | … | 0.39 | 2018-12-13 | … | … |
| … | … | 6.18 | 2001-06-18 | … | … |
| 4XMM J122809.3+435358 | 0.67 | 0.71 | 2018-05-11 | … | … |
| … | … | 7.18 | 2002-05-25 | … | … |
| … | … | 1.17 | 2022-05-08 | … | … |
| … | … | 9.41 | 2002-06-02 | … | … |
| 4XMM J122936.2+075423 | 0.36 | 0.90 | 2016-01-05 | 0.17 | 2010-02-27 |
| … | … | 0.04 | 2004-01-01 | 0.75 | 2015-02-24 |
| … | … | 0.78 | 2016-01-07 | 0.11 | 2019-04-17 |
| … | … | 1.02 | 2016-01-09 | 0.35 | 2020-04-09 |
| 4XMM J125747.0+272919 | … | 0.75 | 2000-06-27 | 2.98 | 2010-11-11 |
| … | … | 2.53 | 2006-06-14 | 1.09 | 2004-02-17 |
| … | … | 0.84 | 2000-06-11 | 0.27 | 2020-03-04 |
| … | … | 0.43 | 2012-06-02 | 0.33 | 2020-03-04 |
| … | … | 3.97 | 2010-12-05 | … | … |
| … | … | 3.69 | 2010-12-11 | … | … |
| … | … | 3.34 | 2010-06-18 | … | … |
| … | … | 3.68 | 2010-06-16 | … | … |
| … | … | 2.69 | 2010-06-20 | … | … |
| … | … | 3.24 | 2010-06-24 | … | … |
| … | … | 4.39 | 2010-12-03 | … | … |
| … | … | 1.98 | 2006-06-11 | … | … |
| … | … | 0.66 | 2012-06-04 | … | … |
| 4XMM J134800.1+263131 | 1.69 | 1.16 | 2003-01-13 | 1.53 | 2012-03-25 |
| … | … | 1.73 | 2004-01-25 | 0.35 | 2021-04-25 |
| … | … | 0.21 | 2000-06-26 | 0.78 | 2014-04-03 |

Table 5
(Continued)

| 4XMM Name | Photo_z[a] | XMM-Newton Flux[b] | Obs. Time | Chandra Flux[c] | Obs. Time |
|---|---|---|---|---|---|
| ⋯ | ⋯ | ⋯ | ⋯ | 0.33 | 2019-04-18 |
| ⋯ | ⋯ | ⋯ | ⋯ | 0.56 | 2013-04-10 |
| ⋯ | ⋯ | ⋯ | ⋯ | 0.63 | 2013-04-08 |
| ⋯ | ⋯ | ⋯ | ⋯ | 0.32 | 2018-04-15 |
| ⋯ | ⋯ | ⋯ | ⋯ | 1.62 | 2016-09-01 |

**Notes.**
[a] Photo-z from SDSS. Only for 4XMM J134800.1+263131; photo-z is from Y. Fu et al. (2024).
[b] XMM-Newton 0.2–2.0 keV fluxes in units of $10^{-14}$ erg s$^{-1}$ cm$^{-2}$.
[c] Chandra 0.2–2.0 keV fluxes in units of $10^{-14}$ erg s$^{-1}$ cm$^{-2}$.

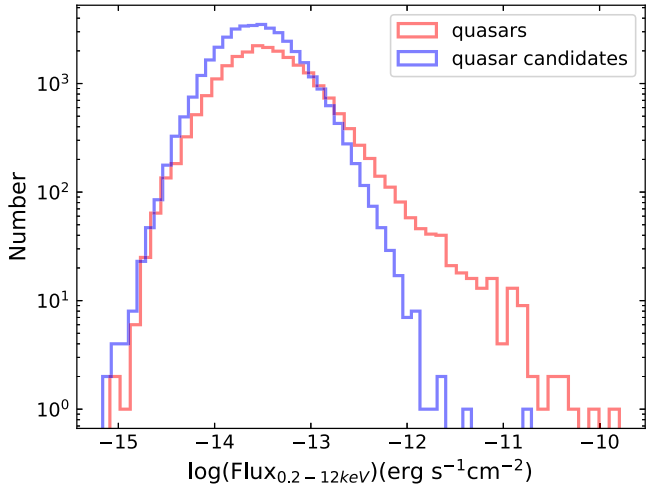

**Figure 4.** Distribution of the broadband X-ray flux for classified quasar candidates and known quasars in training data.

variable AGNs, quasars, or tidal disruption event sources. This step provides independent confirmation of the robustness of our quasar classification and extreme variable quasar selection. For the remaining highly variable quasar candidates, we visually inspect the light curves and corresponding X-ray images individually. Sources with only two observations and large flux uncertainties at both epochs are rejected. Additionally, we exclude sources where variability is due to "problematic" observations, such as those detected near the edges of the field of view, in CCD gaps, or on readout streaks associated with saturated or bright sources.

Finally, the carefully constructed sample comprises 12 quasar candidates, which display extreme variability with a factor greater than 10 in the 0.2–2 keV range between XMM-Newton observations. In Figure 3, we present the soft X-ray light curves for these highly variable quasar candidates. As shown in the figure, seven objects are also detected by the Chandra X-ray Observatory. The presented Chandra soft flux corresponds to the flux in the 0.2–2.0 keV band, as released in the Chandra Source Catalog, assuming a power-law spectral model with a photon index of 2.0 and accounting for Galactic extinction (I. N. Evans et al. 2024).

We also search for the X-ray observations of these quasars from other X-ray satellite missions, such as ROSAT, Swift, and SRG/eROSITA. None of these sources are detected in the RASS. The result can be attributed to the relatively deeper observations in the 4XMM-DR3 catalog compared to those from ROSAT, of which the flux sensitivity is ∼$10^{-13}$ erg cm$^{-2}$ s$^{-1}$ (W. Voges et al. 2000). There are eight quasar candidates located within the range $0° < l < 180°$, while no sources are detected in the X-ray catalog of SRG/eROSITA during the all-sky survey in the $0° < l < 180°$ celestial hemisphere. Additionally, five sources have multiple Swift-X-ray telescope (XRT) observations; however, the measured soft X-ray fluxes are not sufficiently reliable due to significant uncertainties associated with these relatively fainter quasar candidates. As a result, we have excluded these measurements from the Swift-XRT observations. The observation log of these 12 highly variable quasars is reported in Table 5.

## 5. Discussion

We compare the broadband (0.2–12 keV) X-ray flux distributions of the 37,818 quasar candidates with those of the TD in Figure 4. The quasar candidates exhibit a systematically lower X-ray flux compared to the TD quasars. Specifically, at flux below $10^{-13}$ erg s$^{-1}$ cm$^{-2}$, the fraction of quasar candidates is noticeably higher than that of TD quasars (Figure 4). This discrepancy likely reflects the fact that brighter quasars are easier to study and classify using spectroscopy, as discussed in H. Yang et al. (2022). Deep optical spectroscopic observations are necessary to identify and characterize the nature of these faint quasar candidates.

The results above indicate that our selection of highly variable quasar candidates may extend the X-ray highly variable quasar sample into a relatively faint range. We then construct a comparative sample of quasars that exhibit variability of at least a factor of 10, as observed in the XMM-Newton survey. To compile this sample, we select highly variable quasars from the 16,826 quasars in TD using the same criteria outlined in Section 4.2. Approximately 40 quasars meet this selection criterion. As described in Section 4.2, 25 of the initially selected highly variable quasar candidates are identified as classified quasars in the NED database. We visually inspect the light curves of these 65 quasars and eliminate those with significant flux uncertainties. In the end, 43 highly variable quasars remain in the sample.

We show the soft X-ray flux variation versus optical SDSS i-band magnitude for the 43 highly variably quasars and 12 quasar candidates in Figure 5. For each object, we present only the brightest and faintest soft X-ray XMM-Newton flux for clarity. Our selected variable quasar candidates, identified using the RF algorithm, extend the quasar sample with extreme soft X-ray variability to an optical faint end at i-band magnitude of 22. In the figure, we also show the distribution of AGNs with soft X-ray flux drops by a factor $\gtrsim 10$, as reported by D. Bi et al. (2015), who compared the ROSAT and XMM-Newton survey data. As shown in Figure 5, all AGNs/quasars, whether with bright or faint optical emission, exhibit highly variable soft X-ray emission.

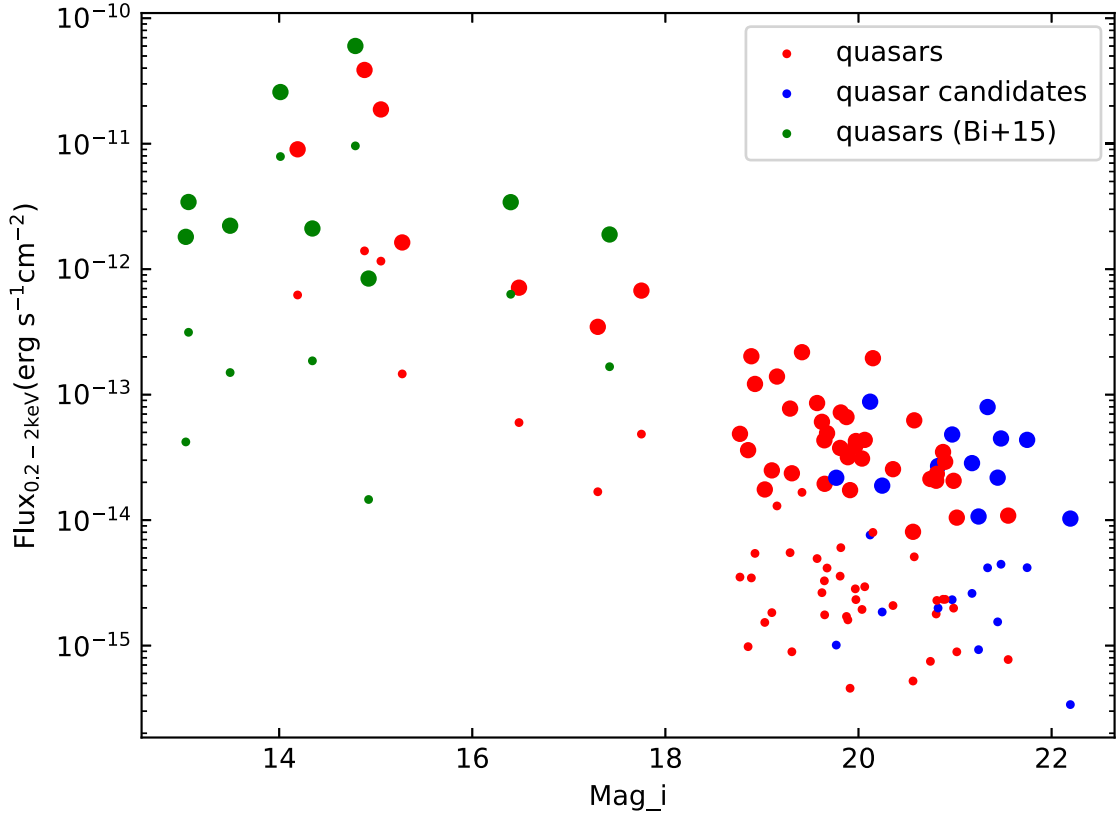

**Figure 5.** Soft X-ray flux vs. SDSS *i*-band magnitude for the 12 highly variable quasar candidates. As a comparison, we also present the distribution of highly variable quasars in training data and in D. Bi et al. (2015). For clarity, only the brightest (larger symbols) and faintest (smaller symbols) soft X-ray states are shown for each object.

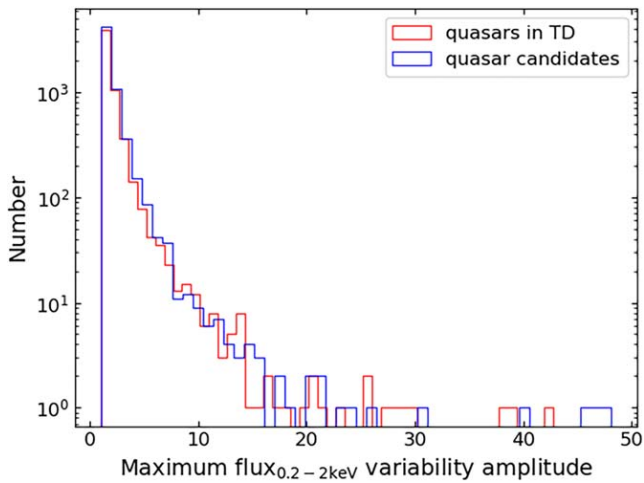

**Figure 6.** Distribution of the maximum flux variability amplitude for classified quasar candidates and known quasars in training data.

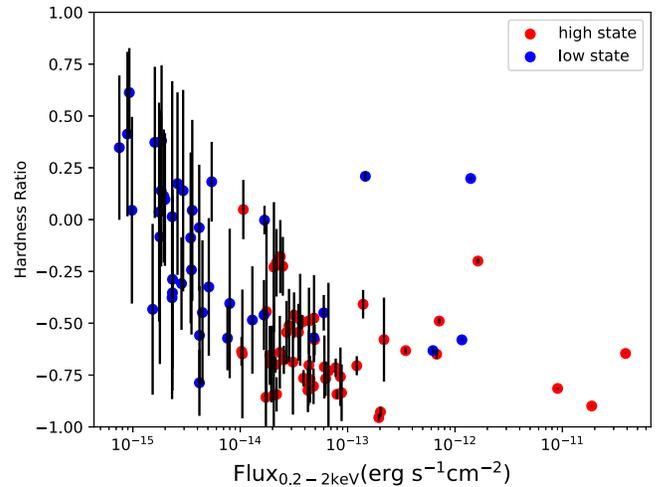

**Figure 7.** Hardness ratio vs. soft X-ray flux for known quasars and quasar candidates with highly variable X-ray emission.

As shown in Figure 5, the soft X-ray fluxes of the quasars brighter than $i \sim 18$ are predominantly distributed in the range of $10^{-11}$–$10^{-13}$ erg s$^{-1}$ cm$^{-2}$. In contrast, quasars with $i \gtrsim 18$ exhibit soft X-ray fluxes between $10^{-13}$ and $10^{-15}$ erg s$^{-1}$ cm$^{-2}$. None of the 12 selected highly variable quasar candidates are detected in the ROSAT survey. This suggests that for optically faint quasars, the soft X-ray fluxes are typically no brighter than the flux limit of the RASS catalog, approximately a few $\times 10^{-13}$ erg s$^{-1}$ cm$^{-2}$. These results also imply that quasars with variability factors of 100 or more are exceedingly rare. The rarity of highly variable quasars is clearly shown in Figure 6, which presents the distribution of the maximum soft X-ray variability flux ratio of quasars in TD and quasar candidates.

A comprehensive analysis of the X-ray and multiwavelength data can help to probe the origin of their extreme X-ray variability. However, the study is out of the scope of the paper. Extreme X-ray flux variations are mostly accompanied by dramatic X-ray spectral shape variations (e.g., Y. Ai et al. 2020; J. Huang et al. 2020). Here we present a brief analysis of the X-ray shape variation with the diagnostics of hardness ratio. The use of X-ray color can provide some initial hints and support our later research. The hardness ratio is defined as $HR = (H - S)/(H + S)$, where H and S are the source counts rate in the 0.2–2 keV (soft) and 2–12 keV (hard). For the 43 highly variably quasars and 12 quasar candidates we estimate the hardness ratio at high and low state with the count rate released in 4XMM-DR13. The observations with $S/N < 1$ at soft band or high band are rejected.

In Figure 7 we present the hardness ratio and corresponding flux at high and low state for the highly variably quasars/quasar candidates. Clearly the X-ray spectrum softens as the flux increases. This "softer when brighter" phenomenon has been well observed in radio-quiet AGNs/quasars (e.g., B. Trakhtenbrot et al. 2017). The result indicates that their X-ray variability might be attributed to changing accretion rate. However, there are many factors that can affect the X-ray spectra of AGNs, such as absorption by warm gas and torus or obscuration by a clumpy disk wind. In such cases the X-ray variability can be explained by changing obscuration. The study of extremely X-ray variable AGNs (e.g., Z. Zhang et al. 2025) suggests that the X-ray variability can be ascribed to changing accretion rate or obscuration. We will perform comprehensive analyses of their multiwavelength properties and investigate the possible origin of their extreme X-ray variability in the following paper.

## 6. Summary

In this study, we apply a random forest algorithm to classify X-ray sources in 4XMM-DR13 using data from the X-ray, optical, and infrared bands, along with the spectral classes of known X-ray sources. A total of 52,486 quasar candidates are classified, primarily located in the relatively faint X-ray region compared to known quasars. From this data set, we select a sample of 12 highly variable quasar candidates within the XMM-Newton survey. All of these candidates exhibit variability in soft X-ray flux by a factor of at least 10. Compared to previous studies, our sample extends the known population of highly variable quasars to the faint end of the optical band, with *i*-band magnitude of 22. Acquiring their rest-frame ultraviolet/optical spectra would help confirm whether they are indeed quasars. A joint investigation of extreme X-ray variability alongside variability in the optical and midinfrared bands could provide valuable insights into the physical nature of these objects. Our study of these highly variable quasars will be further developed in subsequent papers in this series.

## Acknowledgments

We greatly appreciate the referee's insightful suggestions, which have significantly contributed to improving our paper. We acknowledge the support of the Natural Science Foundation of Top Talent of SZTU (GDRC202208), Shenzhen Science and Technology program (JCYJ20230807113910021), and Guangdong Basic and Applied Basic Research Foundation No. 2022A1515012151. We acknowledge the National Natural Science Foundation of China under grant Nos. 12133001 and 12273076.

This research has made use of data obtained from the 4XMM XMM-Newton serendipitous source catalog compiled by the 10 institutes of the XMM-Newton Survey Science Centre selected by ESA. This publication makes use of data products from the Wide-field Infrared Survey Explorer, which is a joint project of the University of California, Los Angeles, and the Jet Propulsion Laboratory/California Institute of Technology, funded by the National Aeronautics and Space Administration.

We acknowledge the LAMOST and SDSS databases. The Guoshoujing Telescope (the Large Sky Area Multi-object Fiber Spectroscopic Telescope (LAMOST)) is a National Major Scientific Project built by the Chinese Academy of Sciences. Funding for the project has been provided by the National Development and Reform Commission. LAMOST is operated and managed by the National Astronomical Observatories, Chinese Academy of Sciences.

Funding for the Sloan Digital Sky Survey V has been provided by the Alfred P. Sloan Foundation, the Heising-Simons Foundation, the National Science Foundation, and the Participating Institutions. SDSS acknowledges support and resources from the Center for High-Performance Computing at the University of Utah. The SDSS website is www.sdss.org. SDSS is managed by the Astrophysical Research Consortium for the Participating Institutions of the SDSS Collaboration, including the Carnegie Institution for Science, Chilean National Time Allocation Committee (CNTAC) ratified researchers, the Gotham Participation Group, Harvard University, Heidelberg University, the Johns Hopkins University, L'Ecole polytechnique fédérale de Lausanne (EPFL), Leibniz-Institut für Astrophysik Potsdam (AIP), Max-Planck-Institut für Astronomie (MPIA Heidelberg), Max-Planck-Institut für Extraterrestrische Physik (MPE), Nanjing University, National Astronomical Observatories of China (NAOC), New Mexico State University, the Ohio State University, Pennsylvania State University, Smithsonian Astrophysical Observatory, Space Telescope Science Institute (STScI), the Stellar Astrophysics Participation Group, Universidad Nacional Autónoma de México, University of Arizona, University of Colorado Boulder, University of Illinois at Urbana-Champaign, University of Toronto, University of Utah, University of Virginia, Yale University, and Yunnan University.

This work has made use of data from the European Space Agency (ESA) mission Gaia (https://www.cosmos.esa.int/gaia), processed by the Gaia Data Processing and Analysis Consortium (DPAC; https://www.cosmos.esa.int/web/gaia/dpac/consortium). Funding for the DPAC has been provided by national institutions, in particular the institutions participating in the Gaia Multilateral Agreement.

This research has made use of the NASA/IPAC Extragalactic Database (NED), which is operated by the California Institute of Technology under contract with the National Aeronautics and Space Administration.

This research has made use of the SIMBAD database, operated at CDS, Strasbourg, France.

## ORCID iDs

Yanli Ai  https://orcid.org/0000-0003-4897-4106
Yanxia Zhang  https://orcid.org/0000-0002-6610-5265
Yuming Fu  https://orcid.org/0000-0002-0759-0504
Liming Dou  https://orcid.org/0000-0002-4757-8622
Xue-Bing Wu  https://orcid.org/0000-0002-7350-6913